# Tracking Fast Neural Adaptation by Globally Adaptive Point Process Estimation for Brain-Machine Interface

Shuhang Chen, Xiang Zhang, Xiang Shen, Yifan Huang, and Yiwen Wang[*], *Senior Member, IEEE*

*Abstract*—Brain-machine interfaces (BMIs) help the disabled restore body functions by translating neural activity into digital commands to control external devices. Neural adaptation, where the brain signals change in response to external stimuli or movements, plays an important role in BMIs. When subjects purely use neural activity to brain-control a prosthesis, some neurons will actively explore a new tuning property to accomplish the movement task. The prediction of this neural tuning property can help subjects adapt more efficiently to brain control and maintain good decoding performance. Existing prediction methods track the slow change of the tuning property in the manual control, which is not suitable for the fast neural adaptation in brain control. In order to identify the active neurons in brain control and track their tuning property changes, we propose a globally adaptive point process method (GaPP) to estimate the neural modulation state from spike trains, decompose the states into the hyper preferred direction and reconstruct the kinematics in a dual-model framework. We implement the method on real data from rats performing a two-lever discrimination task under manual control and brain control. The results show our method successfully predicts the neural modulation state and identifies the neurons that become active in brain control. Compared to existing methods, ours tracks the fast changes of the hyper preferred direction from manual control to brain control more accurately and efficiently and reconstructs the kinematics better and faster.

## I. Introduction

NEURAL adaptation illustrates the change of neural activities in response to behaviors or external stimuli from the single cell level to neural ensemble level while the subject interacts with the environment [1, 2]. For example, in the visual system, some neurons in the V1 cortex have been observed to enhance their responsiveness in an orientation discrimination task [3] or contrast discrimination task [4], and in the primary motor cortex, neurons have been observed to change their modulation depth in resistance muscle training [5]. Importantly, this kind of neural tuning change has also been observed in neuroprosthesis control in brain-machine interface (BMI) systems [6, 7].

BMIs are systems that translate neural activities into digital commands to control external devices [8]. The goal of such systems is to help the disabled restore their motor functions by controlling a neuroprosthesis to accomplish their movement intents. Neural adaptation is critical for the subject to achieve pure brain control on a prosthesis in a closed-loop interaction. This is because the subject needs to adjust their neural firing patterns as the dynamic input to the decoder according to the sensory feedback on the status of the prosthesis [9-11]. Studies have observed that neurons in the motor cortex change their tuning property at the single neuron level when the subject learns to purely use brain signals to move the prosthesis, without actual limb movement, which is called brain control (BC) mode. Specifically, motor neurons were observed to increase their firing rates on the specific directions when human patients learned to control a cursor or a prosthetic limb with their brain signals [5, 12]. There have also been observations of neural adaptation in non-human BMI experiments. In 2011, Ganguly [13] conducted neural recording with macaque monkeys who were trained to move their arms to reach directional targets in a center-out task (manual control (MC) mode). The preferred direction and modulation depth of M1 neurons were observed to change abruptly when the control mode was switched from MC to BC. Later, Orsborn [14] designed a new experiment, where a subset of single neurons were manually selected to participate in BC mode instead of the whole recording. They found the selected neurons started to adapt to the new decoder, changing their tuning property, such as increasing the firing rate on the preferred direction, or shifting the preferred direction in order to accomplish the same BC task.

The above studies suggest that neural adaptation can be driven by external stimuli, such as the switching of tasks, or new design of the decoders. Thus, if the adaptation of the individual neural tuning property can be predicted, it helps researchers to identify the single neurons that are likely to become more active participants in the BC process and to understand how the change of neural tuning property helps BC. Moreover, the prediction of neural tuning provides extra information to enable the decoder to follow the ever-adjusting neural patterns during the closed-loop interaction. As a result,

Research supported by grants from Shenzhen-Hong Kong Innovation Circle (Category D) (No. SGDX2019081623021543), the National Natural Science Foundation of China (No.61836003),. Sponsorship Scheme for Targeted Strategic Partnership (FP902), special research support from Chao Hoi Shuen Foundation.

Shuhang Chen is with the Program of Bioengineering, the Hong Kong University of Science and Technology, Kowloon, Hong Kong (e-mail: schenbx@connect.ust.hk).

Xiang Zhang, Xiang Shen, Yifan Huang and Yiwen Wang are with the department of Electronic and Computer Engineering, the Hong Kong University of Science and Technology, Kowloon, Hong Kong. Yiwen Wang is also with the department of Chemical and Biological Engineering, the Hong Kong University of Science and Technology, and serves as corresponding author (e-mail: eewangyw@ust.hk).

the subject can quickly master BC with better performance.

The point process is an important computational tool to extract the information of a single neural tuning property from discrete spike timing. The neural tuning property can be parameterized by statistical methods [15, 16]. Specifically, the linear-nonlinear-Poisson (LNP) encoding model first defines a preferred hyper-direction obtained by linear projection using the spike-trigger average [17]. Then the preferred hyper-direction is followed by a nonlinear function approximated by the conditional spike firing probability given the stimuli or movements. Finally, the spike timing is modelled as a Poisson process given the firing probability. When the neural tuning parameter changes over time, the distribution of the interval between the adjacent spikes will become different, even for the same stimulus or movement.

Brown et al. [18] were the first to implement the point process to track the dynamic tuning property. They defined a log likelihood of the point process to describe how the spike timing encodes the stimuli. They developed a gradient-based point process filter to track the change of the preferred position from a place cell spike recorded from the hippocampus of rats. Frank [19] tested the gradient-based point process on real data, in which the optimal preferred position at each time instance was searched by a steepest gradient descent method given the rat's real trajectory. Eden [20] and Ergun [21] further developed the point process method using a state-observation model, in which the neural tuning parameters were treated as a state evolving over time. Based on the observation of a discrete spike train, the posterior of tuning parameter state at each time instance were inferred probabilistically by recursive Bayesian estimation. This method was tested on simulation data, and real data recording from place cells in a rat's hippocampus.

In the area of motor BMIs, Wang [22] developed sequential Monte Carlo point process (SMCPP) estimation and implemented it for movement decoding. Instead of estimating the neural tuning parameter, the kinematics is decoded as the state from the observation of multi-channel spike trains in the primary motor cortex, where the posterior of the movement state that nonlinearly relates to neural activity is fully characterized by a set of particles. SMCPP decodes the kinematics given a fixed neural encoding model, which builds the foundation to further track the time-variant tuning parameter given the estimation of movement. This is because the existing point process methods estimate the tuning parameter given the ground truth of the kinematics, while the actual kinematics is not accessible for paralyzed patients.

A dual-model structure has been proposed to estimate the tuning parameter and kinematics at the same time [11, 23, 24]. For example, Dangi and Shanechi [11, 23] combined two Kalman point process filters to decode the trajectory in a directional center-out task and the neural modulation depth in the cosine tuning curve. The better tracking of the modulation depth helped the subject monkey to learn the BC of a target-reach task in a short time. Wang [24] extended the single-SMCPP into a dual-SMCPP (DSMCPP) structure to predict the continuous trajectory of a monkey's hand movement, as well as the nonlinearity in an LNP encoding model. The DSMCPP method successfully tracked the gradual change in nonlinear neural tuning and improved the kinematics decoding performance compared to the static model.

However, the above point process methods assume the neural tuning parameter gradually changes, because the models define the state evolution process which limits the search range in the neighborhood of the previous tuning parameter state. In reality, when subjects start to learn a BC task, some neurons actively engage in exploring all the possibilities of tuning parameters, such as abruptly altering the tuning direction [13, 14] to adjust the decoder output to the desired area. The local search on the tuning parameter can limit the adaptation speed of the decoder, degrade the decoding performance, and therefore slow the learning process.

We are interested in computationally explaining how the brain develops an effective mechanism to adjust neural tuning during the learning process of BC. To identify the neurons that engage in exploring possible preferred directions to accomplish the BC task, here, we propose a globally adaptive point process (GaPP) method to track the neural tuning parameter, specifically the preferred direction, and decode the kinematic state simultaneously from the observation of the spike trains. Compared with the existing methods to track the slow change of tuning depth [24] for a well-trained task, the main difference is that the proposed method conducts a global search on the tuning parameters. Thus, the new method is more suitable for the BC task, which requires fast exploration of the preferred direction. Instead of a direct search on the neural preferred direction in the high dimensional space, we propose to estimate a neural modulation state, which represents how the neurons adjust the firing timing in response to different movements, and then decompose the estimation into the preferred hyper-direction. The neural modulation state is defined as the inner product between the hyper preferred direction and the current kinematics vector, which is the linear projection in the instantaneous linear-nonlinear-Poisson (LNP) encoding model [17]. The larger range that the neural modulation state changes, the more the neuron is related to the kinematics. As the similarity measurement, the neural modulation state has the advantage as a scalar with the finite range within $[0,1]$, which allows the global search over all the possible values. By generating two sets of particles as in sequential Monte Carlo estimation, a probabilistic model is built to estimate the neural modulation state and kinematics state from spike train observation by maximizing their posterior probabilities respectively in the dual-model structure. In particular, the neural modulation states are explored by the particles searching globally within $[0, 1]$ with a small probability. Another set of the particles is responsible to reconstruct the posterior pdf of the movement from the neighborhood of previous estimation guided by the state function. The gradient descent method is implemented to decompose the estimation of the neural modulation states into hyper preferred directions and the reconstruction of the movement.

We validate our method in multiple segments of real data collected from two Sprague Dawley (SD) rats performing a two-lever discrimination task under both MC and BC. We use the proposed method to identify the neurons that become more active in BC mode by estimating the time-variant neural modulation state from the neural spike trains. We further predict the neural preferred direction and reconstruct the kinematics. The results are compared with DSMCPP, which searches the neural tuning parameters in the neighborhood of the previous estimation. Comparing to linear and Gaussian assumptions in [23], DSMCPP and GaPP share the same assumptions on nonlinear evolution and non-Gaussian distribution of states [24]. A Kolmogorov-Smirnov statistical test (KS-plot) is used to qualify how well the time variant neural preferred direction is predicted. And the performance of kinematics reconstruction is evaluated by normalized mean square error (NMSE), which is expected to be less when the tuning direction is properly predicted.

The rest of this paper is organized as follows: Data acquisition and the behavior tasks are described in Section II.A. Section II.B shows the details of the GaPP algorithm to track the neural modulation state, followed by the decomposition of the preferred direction and kinematics reconstruction in Section II.C and the evaluation methods in Section II.D. Section III shows the results on *in vivo* neural data and comparisons with existing methods.

## II. DATA COLLECTION AND METHODOLOGY

### A. Data Collection

The motor BMI experiment paradigm on rats was designed and implemented at The Hong Kong University of Science and Technology (HKUST). The animal handling procedures were approved by the Animal Ethics Committee at HKUST. The rats were trained to perform a two-lever discrimination task. At the beginning of each trial, an audio tone with frequency of 10 kHz or 1.5 kHz, was randomly presented. The rats were required to distinguish the audio tone, and press the corresponding lever (high tone corresponding to high lever, and low tone corresponding to low lever). The lever should be held over 500 ms to get water reward. Audio feedback with the same frequency would be presented to indicate success after the 500-ms holding. The try-out time was 6 s for each trial. Any early release, no pressing, or wrong pressing would lead to no reward. And the inter-trial interval was set randomly from 3 to 6 s. In the MC mode, the rats used their right forepaw to press the levers, while in the BC mode, the rats needed to adjust their neural activity through an online decoder to output the action of virtually pressing the high or low lever without actual limb movement [25].

A 16-channel microelectrode array was implanted in the left primary motor cortex (M1), which controls the right forepaw. A multi-channel acquisition processor Plexon (Plexon Inc., Dallas, Texas) was applied to record the extracellular potential with a sample rate of 40 kHz while the subjects were performing the task. The recorded potential was filtered with a 4-pole Butterworth high-pass filter at the threshold frequency of 500 Hz. The filtered potential was detected by a threshold within $-5\sigma \sim -3\sigma$ ($\sigma$ is the standard deviation of potential amplitude). The detected spikes of a single channel were further sorted into one or more units based on the shape of waveforms. The spike timing of individual units was recorded.

The corresponding behavior events were recorded by a behavioral chamber (Lafayette Instrument, USA) synchronized with the neural recording through the digital ports, and down-sampled to 100 Hz as spike trains. In MC mode, the neural spike trains and corresponding kinematics of successful trials were segmented as 500 ms before and after the cue onset as well as the press onset, and 200 ms after success. The movements were labelled as $[0,0]$ for the rest state (before audio cue) and $[1,1]$ and $[1,-1]$ for high-lever holding and low-lever holding, respectively. For the well-trained MC session of Rat A, a total 44 successful high-lever trials and 56 successful low-lever trials were conducted, while for Rat B, there were 102 successful high-lever trials and 105 successful low-lever trials. A Kalman decoder was built to predict the 2-dimensional movement states, which were smoothed with a sigmoid function in the training set from the neural firing rate observation function [25, 26].

In BC mode, the neural activities were continuously interpreted online into the 2D movement state every 100 ms using the Kalman decoder trained by well-trained MC data. Due to the limited performance of the decoder and the different conditions between MC and BC, the subjects had to explore their neural preferred direction and adjust their neural activity to achieve the BC task. The virtual levers were determined to have been pressed when the brain states stayed within the desired region in the 2D space over 500 ms for the high lever and the low lever respectively. A total 68 successful high-lever trials and 98 successful low-lever trials in well-trained BC session of Rat A, and 102 high-lever trials and 101 low-lever trials for Rat B were conducted. For point process analysis, the spike timing information was binned by a 10-ms time window for each neuron to build the spike train observation. Within these intervals, 92% of those with spikes had only one spike. For each time window, 1 was assigned to intervals with one or more spikes; otherwise, 0 was assigned. In total, 17000 data points under the MC and 150000 data points under the BC of Rat A were used for the training and testing of the neural tuning prediction. For Rat B, 20000 MC data points and 70000 BC data points were used for the same purpose. These data points are divided into 15 segments covering MC and BC to validate the statistical performance. Each segment containing over 20000 data points cascade the MC and BC data to test whether the decoder can effectively track the large changes of tuning property in short time.

### B. Tracking the Neural Modulation State by Globally Adaptive Point Process Estimation

We propose to predict a neural modulation state for the individual neuron by global search from point process observation, and then decompose the modulation state into the neural preferred direction and kinematics reconstruction. The neural modulation state is defined to describe how the neural spike trains modulate the current high-dimensional kinematics at time $t$. For a neuron $n$, the neural modulation state $z_t^n$ is the inner product of the kinematics vector and the hyper tuning preferred direction, expressed as

$$z_t^n = (K_t^n)^T x_t, \quad (1)$$

where $x_t$ is the high-dimensional kinematics vector at time $t$. Here, we define it as $x_t = [p_x, p_y, v_x, v_y, 1]^T$ for our data, which includes the position and the velocity in a 2D-plane with a bias. The linear filter $K_t^n$ represents the $n^{th}$ neuronal tuning preferred direction in the hyperspace, which could change over time and can be estimated by spike-trigger regression if given the ground truth kinematics [17, 24]. $K_t^n$ points to the direction in which the neuron has the highest firing probability, and its length presents the modulation depth of the firing property. Therefore, the neural modulation state $z_t^n$ actually describes the similarity between the current kinematics vector and hyper-tuning preferred direction over the time course. Note that $z_t^n$ has finite range[0,1], because it is a linear estimation of the firing probability which is able to be searched globally. It connects the kinematics to the neural spike train through the instantaneous linear-nonlinear-Poisson model [17] as follows:

$$\lambda^n(t|z_t^n) = \lim_{\Delta t \to 0} \frac{\Pr(N_{t+\Delta t}^n - N_t^n = 1|z_t^n, N_t^n)}{\Delta t} \quad (2)$$
$$= \exp(a^n z_t^n + b^n),$$

where $\lambda^n(t|z_t^n)$ is the instantaneous conditional firing probability of neuron $n$ at time $t$, and $a^n$ and $b^n$ are two parameters for exponential fitting of the nonlinear function for neuron $n$ which are estimated and fixed from the training data. The number of spikes $N_t^n$ up to time $t$ is modeled by an inhomogeneous Poisson model with the instantaneous conditional intensity function $\lambda^n$:

$$\Pr(\Delta N_t^n | z_t^n) = \frac{(\lambda^n(t|z_t^n)\Delta t)^{\Delta N_t^n}}{\Delta N_t^n!} \exp(-\lambda^n(t|z_t^n)\Delta t), \quad (3)$$

where $\Delta N_t^n = N_{t+\Delta t}^n - N_t^n$ represents the number of spike events within time interval $\Delta t$.

We propose a GaPP method to estimate the time-variant neural modulation state from the spike train observation. Since the neural modulation state is linearly relative to the hyper-preferred direction, regression models can be implemented to decouple the preferred direction from the estimated neural modulation with given kinematics. Different from the DSMCPP [24] that models the state with the slow transition function, here, we allow $z_t^n$ to be globally explored with a small probability, which is suitable for switching from MC to BC. Here, we switch from a continuous time course to a discrete time framework to present our method. The observation interval is partitioned into $\{t_k\}_{k=0}^K$ and the individual interval is represented by $\Delta t_k = t_k - t_{k-1}$. The lower index $k$ will be used to indicate the time step. Therefore, the transition of $z_k^n$ from $z_{k-1}^n$ can be written as

$$\begin{cases} \phi_k^n = F^n z_{k-1}^n + r_k^n & \text{with probability } (1-\psi) & (4a) \\ \gamma_k^n \propto g(\cdot) & \text{with probability } \psi, & (4b) \end{cases}$$

where $\phi_k^n$ represents the possible neural modulation state of neuron $n$ that changes slowly over time, $F^n$ is the transition matrix trained from 10000 training data points by the least squares method, $r_k^n$ is the zero-mean Gaussian noise of neuron n with the covariance $R$ estimated by the residue of linear approximation, $\gamma_k^n$ represents the exploration of all possible neural modulation states of neuron $n$ that the subject adjusts during the adaption to BC, and $g(\cdot)$ generates the possible neural modulation state globally and usually is set as a uniform distribution. Due to the nonlinearity of the tuning function, we propose to generate two sets of particles to derive the neural modulation state both globally and locally from the spike train observation by sequential Monte Carlo estimation. For neuron $n$, two sets of particles, $\{\phi_k^{n,i}\}_{i=1}^{N_s}$ and $\{\gamma_k^{n,i}\}_{i=1}^{N_s}$, are generated at time index $k$ to approximate the posterior density of the neural modulation state. The number of particles in each set is $N_s$ and $i$ is the particle index. $\{\phi_k^{n,i}\}_{i=1}^{N_s}$ are generated in the neighborhood of the previous neural modulation state with a probability $(1-\psi)$. The other set of particles $\{\gamma_k^{n,i}\}_{i=1}^{N_s}$ are distributed uniformly within the finite range [0,1] with a probability $\psi$. The value of $\psi$ indicates how rarely abrupt changes happen in this case and balance local search and global search. It is generally set to a very small value because the exploration of the tuning parameter does not happen very often. Therefore, the local search dominates the posterior density of the neural modulation state at most time indexes. But when abrupt neural adaptation occurs, the global search can drag the posterior out of the local area. The posterior of the neural modulation state $p(z_k^n|\Delta N_k^n)$ is formed by the weighted particles, as in Parzen window estimation [27], from the two sets of particles, locally and globally, as

$$p(\phi_k^n|\Delta N_k^n) = \sum_{i=1}^{N_s} \frac{1}{N_s} w_{k,1}^{n,i} \kappa(\phi_k^n - \phi_k^{n,i}) \quad (5)$$

$$p(\gamma_k^n|\Delta N_k^n) = \sum_{i=1}^{N_s} \frac{1}{N_s} w_{k,2}^{n,i} \kappa(\gamma_k^n - \gamma_k^{n,i}), \quad (6)$$

where $p(\phi_k^n|\Delta N_k^n)$ and $p(\gamma_k^n|\Delta N_k^n)$ are the posterior distribution of the neural modulation state by local and global estimation, respectively, $\kappa(\cdot)$ is a Gaussian kernel, whose size is determined by Silverman's rule[27]. For each particle, the associated weights are initialed equally and updated by the spike train observation as

$$w_{k,1}^{n,i} \propto p(\Delta N_k^n | \phi_k^{n,i}) \quad (7)$$
$$w_{k,2}^{n,i} \propto p(\Delta N_k^n | \gamma_k^{n,i}), \quad (8)$$

where $p(\Delta N_k^n | \phi_k^{n,i})$ and $p(\Delta N_k^n | \gamma_k^{n,i})$ are the likelihood of generating $\Delta N_k$ spikes given tuning functions and tuning parameter estimation, as defined in Eq. (2) and (3). The posterior pdf of the current neural modulation state is then constructed by combining the two sets of particles with their corresponding weights $\{(1-\psi)w_{k,1}^{n,i}\}_{i=1}^{N_s}$ and $\{\psi w_{k,2}^{n,i}\}_{i=1}^{N_s}$. $N_s$ new particles $\{z_k^{n,i}\}_{i=1}^{N_s}$ are resampled from the $2N_s$ particles $\{\phi_k^{n,i}\}_{i=1}^{N_s}$ and $\{\gamma_k^{n,i}\}_{i=1}^{N_s}$ to equalize their new weights [22, 24, 28]. And the posterior density is finalized as

$$p(z_k^n|N_k^n) = \sum_{i=1}^{N_s} \frac{1}{N_s} \kappa(z_k^n - z_k^{n,i}). \quad (9)$$

The estimation of the neural modulation state $z_k^n$ of neuron $n$ uses the expectation of posterior density.

*C. Decomposition on Neural Hyper-Tuning Preferred Direction and Kinematics Reconstruction*

Due to the linear form of the relationship between the neural modulation state and the preferred hyper-tuning direction, gradient descent optimization is here implemented to estimate the preferred hyper-tuning direction $K_k^n$ for neuron $n$ at time index $k$. The mean square error is used as the cost function in the gradient descent:

$$J^n = \frac{1}{2T}\sum_{i=0}^{T}(\tilde{z}_i^n - K^T x_i^*)^2, \quad (10)$$

where $T$ is the history length, which should be long enough to statistically support the estimation of the tuning parameter at time index $k$; $i$ is the history index; $\tilde{z}_k^n$ is the estimated neural modulation parameter at time $k$; and $x_k^*$ is the decoded kinematics vector at time $k$, for which the approach is described later. The update of the preferred hyper-tuning direction uses stochastic gradient descent, and is expressed as

$$\begin{aligned}K_k^n &= K_{k-1}^n - \varepsilon^n \frac{\partial J^n}{\partial K}\bigg|_{K=K_{k-1}^n} \\ &= K_{k-1}^n - \varepsilon^n \frac{1}{T}\sum_{i=0}^{T}(\tilde{z}_{k-i}^n - K^T x_{k-i}^*)x_{k-i}^*\bigg|_{K=K_{k-1}^n},\end{aligned} \quad (11)$$

where $\varepsilon^n$ is the learning rate for neuron $n$. For each updating time index, the cost function considers the past 10000 data points, and the update will not stop until the change of the cost value is smaller than a predefined threshold. Here, we set it as $5 \times 10^{-9}$. Every estimated hyper-tuning preferred direction will be considered static within a short duration and support the estimation on the kinematics from the spike observation at the next time index, which is similar to a dual structure. We use sequential Monte Carlo estimation (SMCPP) [22, 24] to decode the kinematics with predicted tuning parameter in both GaPP and DSMCPP.

*D. Evaluation of Neural Preferred Direction Prediction and Kinematics Decoding Performance*

The Kolmogorov-Smirnov (KS) statistic is used to evaluate the prediction of the neural preferred direction. We feed the estimated neural modulation state into the LNP model, and generate the prediction on the neural firing probability. The goodness-of-fit between the spike train and predicted neural firing activity can be measured by KS plot [27, 29]. Based on the time rescaling theorem, the inter spike intervals conditioned on the firing probability can be rescaled into a uniformly distributed variable. If the estimated conditional firing probability is more accurate, the cumulative distribution of the rescaled variable should stay closer to the 45° line.

We also use information theoretical analysis to validate whether our method identifies the neurons that become more engaged in BC mode. Mutual information is applied to measure the information amount indicating how the spike train modulates the neural modulation state [30, 31]:

$$\begin{aligned}I(\Delta N_{1:T}^n, z^n) = \sum_{z^n} p(z^n) \sum_{\Delta N=0,1} p(\Delta N^n|z^n) \\ * \log_2(\frac{p(\Delta N^n|z^n)}{p(\Delta N^n)}),\end{aligned} \quad (12)$$

where $p(z^n)$ is the pdf approximated by the samples within all the time courses, $p(\Delta N^n|z^n)$ is the conditional firing probability derived from the tuning function of the instantaneous LNP model, and $p(\Delta N^n = 1)$ is calculated as the percentage of the spike count during the entire spike train. We compare the reconstructed neural modulation state estimated from the spike trains and ground truth obtained from both the spike trains and true kinematics, and predict the mutual information using the estimated neural modulation states over time. The neurons with a higher amount of mutual information are identified as the important neurons in BC.

The kinematics decoding performance is evaluated by the normalized mean square error (NMSE) between the true kinematics and the decoded kinematics. NMSE has been widely used in movement decoding work of BMI [10, 14, 23, 24]. This criterion is chosen because two position dimensions both participate in the BC task. The improvement of the kinematics decoding is expected if we can track the time-variant tuning parameter well. The estimation of the neural preferred direction and the kinematics reconstruction by the proposed method are compared with the existing DSMCPP method. And the Student's t-test is conducted to evaluate whether the proposed method improves the performance statistically, over multiple segments of data. We also compare the time spent to achieve 90% of the best decoding performance to judge how fast the algorithm tracks the tuning property.

III. RESULTS

In this section, we implement the proposed GaPP model to estimate the neural modulation state and decompose it into the neural tuning hyper-preferred direction and kinematics reconstruction over multiple segments of real data. The goal is to illustrate if our method can identify the neurons actively involved in BC, to determine how efficient our algorithm is to track the abrupt changes in the hyper-tuning parameter in the transition from MC to BC, and to support better kinematics reconstruction.

Fig. 1 gives an example of the kinematics and corresponding neural spiking raster graph while Rat B is performing the task in MC mode (green background), covering five successful trials, and in BC mode (pink background), also covering five successful trials. Fig. 1(b) and (c) are the horizontal and vertical dimensions of the 2-dimensional position, respectively, where the x-axis is the time in seconds. In MC mode, the movement events are labeled as the resting (yellow bar), press (blue bar) and reward (red bar) states. In BC mode (pink background), the 2-dimensional position is online decoded from multi-unit activity by the Kalman filter. The reward is determined if the decoded trajectory reaches and stays at the desired area for 500 ms, as indicated in Fig 1(a) (cyan area for virtual high lever press, purple area for virtual high lever press). And the yellow area in Fig. 1(a) represents the resting state where the rats need to adjust their brain state and stay in the area for 1.5 s to trigger the next cue. Fig. 1 (d) shows the spike trains of two selected neurons. We can see that neuron 1 fires more spikes in the resting state and generates few spikes in the press state, both in MC and BC mode. For neuron 16, the spike rates of the resting and press states have no obvious difference in MC mode. But in BC mode, neuron 16 generates more spikes in the press state for a high lever and inhibits spike generation in the low-lever

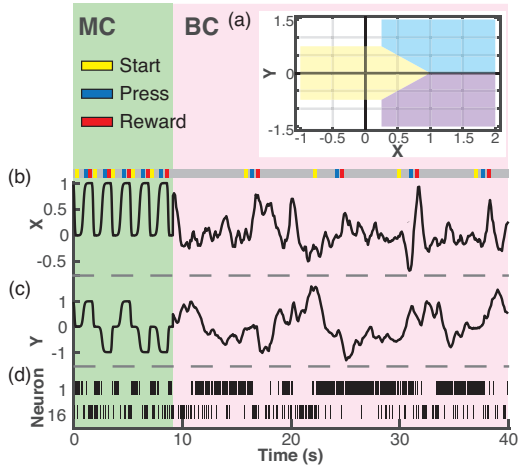

Fig. 1. Real data of kinematics and neural signals. The kinematics includes 2-dimensional position X and position Y in (b) and (c). The MC position (green area) is labeled manually according to the recorded lever press state, where [0, 0] represents the resting state (labelled as yellow bar on the top grey plot in (b) and (c)) and [1, 1] and [1, -1] represent the high-lever press state and low-lever press states, respectively (labelled as blue bar). In BC mode (pink area), the 2-dimensional position is online decoded from multi-unit activities by the Kalman filter. The reward is determined if the decoded trajectory reaches and stays at the desired area (cyan area for high-lever press, and purple area for low-lever press) for 500 ms, as indicated in (a). (d) shows recorded neurons that are strongly related to the BC task.

press state.

For the neural tuning analysis, we use a 100-s sliding-window with 98% overlap to derive the time sequence of the ground truth tuning parameter in the LNP model for MC and BC, respectively. 100-s window is selected because it contains about ten trials which is statistically sufficient to analyze the state of tuning property. We find that the nonlinearity in the tuning function remains similar across MC and BC for most neurons. Some neurons slightly change their hyper-tuning preferred directions $K_k^n$ within one task and have a significant change between two tasks. Therefore, we fix nonlinear function parameters $a^n$ and $b^n$ in the encoding model described in Eq. (2), and only predict the neural modulation state $z_k^n$ and the hyper-tuning preferred direction $K_k^n$ with GaPP. We initialize the neural modulation state at the beginning of the test using the ground truth value. The search range of the modulation state is within 0 and 1, and the prior probability of abrupt change $\psi$ is set as 0.08 after exploring from 0.001 to 0.10. The transition matrix $F^n$ and noise $r_k^n$ are trained from 2000 MC data samples by the least squares method. As a comparison, the neural modulation state estimated by DSMCPP is the product of the estimated kinematics and estimated preferred direction. In the kinematics estimation, the parameters of the transition function, $A$ and $q_k$, are trained from 10000 data samples across the MC and BC modes. Both GaPP and DSMCPP are performed for 10 Monte Carlo runs [22, 24], and the average of the estimations is used to evaluate the decoding performance.

Fig. 2 shows the validation on the estimated neural modulation states of two neurons in two rats. The upper row is for neuron 8 in Rat A, and the lower row is for neuron 16 in Rat B. We feed the estimated neural modulation states, obtained by the proposed GaPP model (black line in Fig. 2(a) and (d)), into the exponential tuning function described in Eq. (2), and generate the neural firing probability (black line in Fig 2 (b) and (e)) over time. The estimated neural firing probabilities are compared with the ground truth (red line in Fig. 2(b) and (e)), which is obtained by smoothing the recorded spike timing with a Gaussian kernel with a 600-ms kernel size.

We also implement DSMCPP to estimate the hyper-tuning preferred direction and kinematics. Then the estimation of the neural modulation state (blue line in Fig. 2(a) and (d)) is derived using Eq. (1) and fed into the same tuning function to generate the firing probability (blue line in Fig. 2 (b) and (e)) for comparison. The green bar is the time separation between the MC and the BC sessions.

We can see from Fig. 2 that GaPP traces the ground truth neural modulation state better than DSMCPP from the switch point of tasks, because GaPP searches the optimal neural modulation state from all the possibilities. We can see GaPP does not track the neural modulation state well when the ground truth neural modulation state is near to zero. This is because when the local search covers the negative range of neural modulation, the likelihood is closer to zero. Therefore, the difference is eliminated by the exponential tuning function, where the neural modulation state is further converted into the firing probability. These advantages result in the better prediction of the neural firing probability. Comparing results for neuron 8 of Rat A, the estimation of the neural modulation state obtained by DSMCPP is further from the ground truth for the whole BC session. For neuron 16 of Rat B, the neural modulation state obtained by DSMCPP is relatively flat at the beginning of the BC session, and then becomes closer to the ground truth firing probability at the end, but still does not track the peaks of the firing probability. This indicates that GaPP can track the neural adaptation of the firing probability faster than DSMCPP, and it is possible that DSMCPP will eventually fail to track.

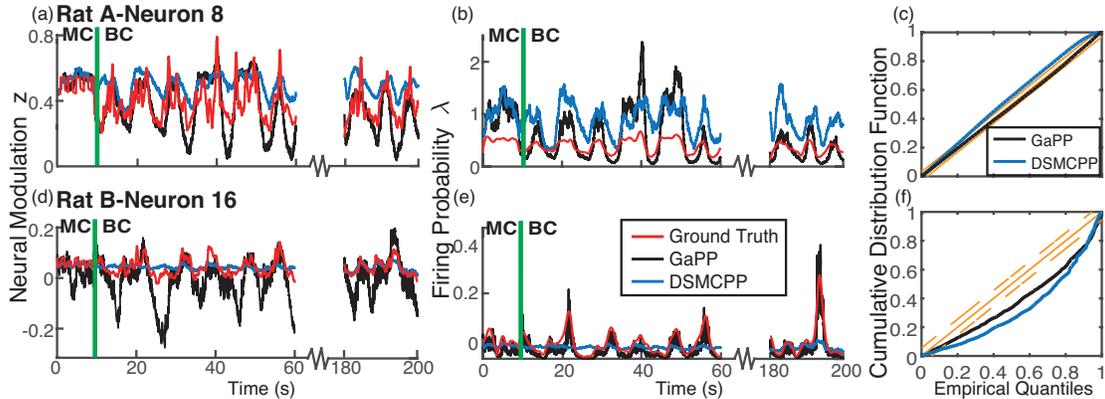

Fig. 2. Validation of estimation of the neural modulation states. The neural modulation state is estimated by the proposed globally adaptive point process model (black), and then serves as the input to the exponential tuning function to generate the firing probability. It is compared with the ground truth neural recording (red). The upper row is for neuron 8 in Rat A, and the lower row is for neuron 16 in Rat B. The x-axis is the time in seconds. The y-axis is the neural modulation state in (a) and (d) and firing probability in (b) and (e). We also implement DSMCPP to estimate the hyper-tuning preferred direction and kinematics, and then derive the neural modulation states and firing probability (blue lines) for the same encoding validation. (c) and (f) give the KS-plot of two neurons for GaPP (black line) and DSMCPP (blue line). The 95% confidence intervals for the KS statistics are plotted by the orange dashed line.

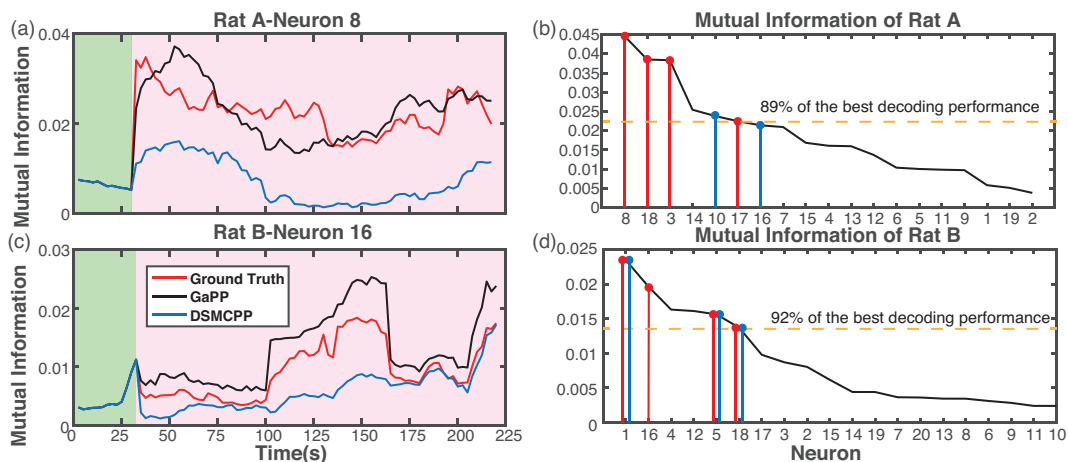

Fig. 3. Identifying the important neurons in BC by estimated mutual information using neural modulation states. (a) and (c) show the comparison of estimated mutual information crossing time for two neurons from two rats. The x-axis is time in seconds and the y-axis is the value of MI. The green and pink backgrounds represent MC and BC respectively. The red line represents the ground truth MI obtained by the observation and real kinematics, the black line is the MI obtained by the neural modulation state estimated by GaPP, and the blue line is the MI conducted by the reconstructed kinematics and predicted preferred direction in DSMCPP. (b) and (d) give the rank of MI in BC mode for two rats, where MI is obtained by GaPP for all the neurons. The neurons labelled by red points match the most important neurons using the ground truth mutual information and the neurons labelled by blue points are the important neurons by DSMCPP. The orange line represents the 89% and 92% of the best decoding performance that is contributed by the top six neurons identified by the GaPP model for the two rats respectively.

Fig 2(c) and (f) show the KS-plot evaluation for the neural modulation state in the BC session. The cumulative distribution of the transformed ISIs using the firing probability estimated by GaPP (black line) is closer to the 45-degree line than that estimated by DSMCPP (blue line), which further validates the better prediction of the neural modulation state by the GaPP model.

The predicted neural modulation states can be used to identify the neurons that actively participate in the BC task. Here we calculate the mutual information using the estimated neural modulation states and the recorded spike train described in Eq. (12). Fig.3 (a) and (c) give the mutual information estimated from the neural modulation states for two neurons, one each from Rat A and Rat B respectively. The information amount measures how the spike train is related to the movements. The x-axis is the time in seconds and the y-axis is the mutual information value. Here we use a 50-s sliding time window with a 48-s overlap to smooth the value of the mutual information over the time covering MC mode (green background) and BC mode (pink background). The red line is the ground truth mutual information obtained from the true kinematics and spike recording, and the black line and blue line are the mutual information obained by GaPP and DSMCPP respectively. We can see that both neurons have a significant increase in their information amount from MC and BC, which indicates the neurons participate more in BC than MC. Notice that the shape of the MI curve in the BC mode is similar to the ground truth, with a correlation coeficient (CC) 0.6944 for neuron 8 of Rat A and 0.8929 for neuron 16 of Rat B. In comparison, the MI curve estimated by DSMCPP has a CC of 0.6434 for neuron 8 of Rat A and 0.7945 for neuron 16 of Rat B. In some parts of Fig. 3 (a) and (c), the mutual information obtained by GaPP is higher than the ground truth. This is because GaPP assumes the neurons fully participate in encoding the kinematics, while they may also contribute to other brain states [32]. GaPP is a more efficient method than DSMCPP in identifying important neurons that engae in BC. This is because GaPP has the advantage of using the directly estimated neural modulation state to compute the mutual information, while DSCMPP has to estimate the hyper-tuning parameter and kinematics respectively and then derive the mutual information value, which increases the compuation complexity.

We select the important neuron subset based on the mutual information using the estimated neural modulation states. The mutual information values of all neurons of Rat A and Rat B in BC are plotted in descending order in Fig. 3(b) and (d) respectively, where the x-axis is the neurons ranked by the mutual information values in descending order and the y-axis is the value of the mutual information. The top six neurons selected by GaPP achieve 89% of the decoding performance conducted by all the recordings for Rat A. For Rat B, meanwhile, the top six neurons selected by GaPP achieve 92% of the decoding performance conducted by all the recordings. For both rats, the neurons labelled by red points match the most important neurons identified by the ground truth mutual information. Meanwhile, the neurons labelled by blue points are the important neurons matched by DSMCPP. This indicates that the neural modulation state estimated by GaPP successfully identifies the important neurons in BC mode and outperform than DSMCPP.

We then decompose the estimated neural modulation states into the hyper-preferred direction as well as the kinematics reconstruction. The initial preferred direction and kinematics are the ground truth value. The learning rate for two example neurons are set as $\alpha = [0.003, 0.01]$ after parameter exploration. For every 200 intervals, historical samples up to 100 s are input to the gradient descent method as one batch. The update does not stop until the change of cost value is less than $5 \times 10^{-9}$. The parameters of preferred direction transition function in DSMCPP are trained from 100-s MC data. The length of history and update frequency is the same as the values in neural tuning analysis of ground truth.

Fig. 4 shows results of tracking the time-varying neural tuning preferred direction for two neurons in Rat A and Rat B. Fig. 4(a) demonstrates the results in the time course. Each row represents one neuron in one rat, and the two columns represent the x and y coordinates of the preferred direction respectively. The x-axis is time in seconds, and the y-axis is the value of one dimension of the preferred direction. The true value of $K$ is plotted by the red line, while the results of GaPP and DSMCPP are drawn by the black line and the blue line, respectively. We can see that the preferred direction estimated by GaPP can catch the abrupt changes from MC to BC with fewer time steps than the estimation by DSMCPP. This demonstrates that GaPP can track the abrupt neural adaption faster by searching the optimal neural modulation state in the full range. In comparison, for neuron 8 of Rat A, the local search in DSMCPP fails to track the significant changes in BC. And in the case of neuron 16 in Rat B, DSMCPP requires a longer time to accumulate the changes made by local search. Meanwhile, the results show that the prediction of the second dimension ($K_2$) always outperforms that of the first dimension ($K_1$), indicating that the second dimension may be preferential over the first dimension. This is because the Y position can not only distinguish between a high lever press (around [1,1] in Fig. 1 (a)) and a low lever press (around $[1, -1]$), but also contributes to determining whether the rat is in the resting or press state, while the X position only contains the information of whether there has been a press or not. Therefore, the proposed method tracks the states with more information with more priority.

The performance of estimating the neural preferred direction can also be described in a 2D-plane, as shown in Fig. 4 (b) where the top and bottom rows show the 2D preferred directions at different time indexes for neuron 8 of Rat A and neuron 16 of Rat B, respectively. The x-axis is the first dimension of the hyper-preferred direction, corresponding to the position in X, and the y-axis is the second one, corresponding to the position in Y. The dashed purple arrow represents the initial MC preferred direction, and the solid red arrow is the ground truth direction in BC at the labelled time. The estimations of GaPP and DSMCPP are plotted by the black and blue arrows respectively. The first column is the initial time of MC when the estimations of the two approaches are both the same as the ground truth. The second column shows the early stage of BC right after the task switches. We can see that the two neurons show obvious differences between the MC (purple dashed arrow) and BC (red solid arrow) parameters, whether in length or direction. At this time, GaPP has started to change to tracking the true preferred direction of BC mode, while the predictions of DSMCPP are still around the MC arrow. The third column shows the time after around five trials in BC mode. At this stage the preferred direction obtained by DSMCPP is still at the initial state in neuron 8 of Rat A, while the estimation of Rat B-neuron 16 starts to leave the MC state and move toward the real BC direction. Meanwhile, the prediction from GaPP has reached the side representing the real direction. This figures in Fig. 4(b) in the instantaneous 2D-plane demonstrate that GaPP reacts to the abrupt neural adaptation more quickly and make a more accurate estimation than DSMCPP, both in terms of modulation depth and direction.

The prediction of the neural tuning parameter will help the reconstruction of the kinematics. A total 10000 points of brain control data are used to train the kinematics evolution model. The tuning parameter is updated every 200 ms by GaPP and the parameter estimator in DSMCPP. The kinematics are decoded using the identified neurons and the important and static tuning neurons. Fig. 5 shows one segment of the kinematics reconstruction at the beginning of BC when switching from MC for Rat B. The two rows are the horizontal and vertical positions in the y-axis, and the x-axis is time in seconds. The ground truth, GaPP result and DSCMPP result are labelled by the red, black

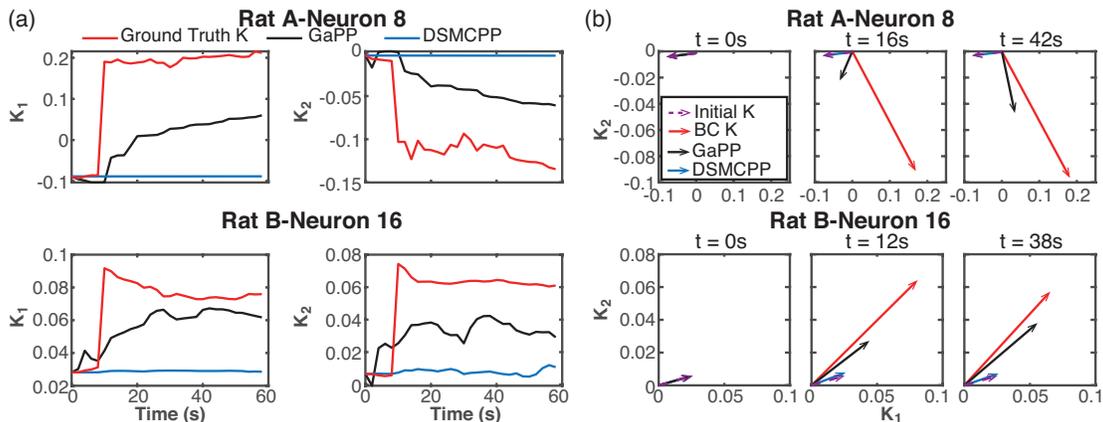

Fig. 4. Estimation of the tuning preferred direction. (a) shows the estimation of preferred direction across time. The x-axis is time in seconds, and the y-axis is the value of the preferred direction. The first column is the first dimension corresponding to the position X and the second column corresponds to the position Y. (b) is the estimation in 2D at the beginning (first column), middle (second column) and end stage (third column) of the tracking.

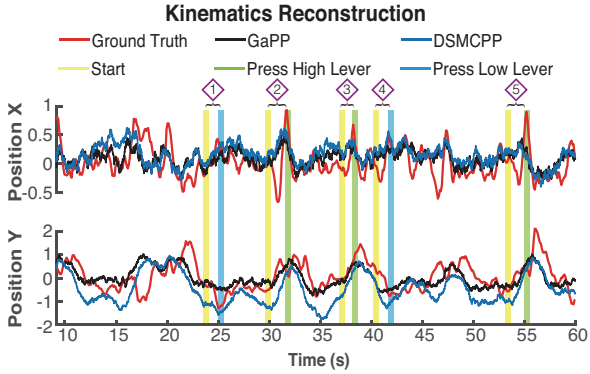

Fig. 5. Kinematics reconstruction with dynamic tuning parameter estimated by the two algorithms. The red line is the ground position, and the black line and blue line represent the reconstruction obtained by GaPP and DSMCPP respectively. The vertical bars label the event as start (yellow bar), press high lever (green bar) and press low lever (cyan bar)

and blue lines respectively. We can see that, in the X-position, the two methods have no apparent difference, and their MSEs are 0.0734 (GaPP) and 0.0901 (DSMCPP) when compared with the ground truth movement. However, in the Y-position, GaPP has a better reconstruction, with a 0.3065 MSE compared to DSMCPP with 0.6405. In particular, at the valleys, DSMCPP always overshoots to a negative amplitude compared to the ground truth movement. This is because the preferred direction estimated by DSMCPP is much shorter than the ground truth preferred direction.

The real movement events of five trials are also labelled in Fig. 5. The yellow bar is the start of trial, green bar is pressing a high lever and blue bar is pressing a low lever. Among these five trials, the two methods can reach the state of pressing a high lever in the high cue trials (2, 3, 5). However, at the start of trial, DSMCPP always reaches too low and never stays long enough in the resting state. This results in the failure to trigger the start cue in BC mode for all five trials. In contrast, GaPP can stay at the rest state and successfully trigger the start for the next trial for continuous BC decoding.

We further test the kinematic reconstruction when the task is switched from MC to BC over multiple segments of data. Fig. 6 illustrates the statistical validation over 15 task switches using the two methods for two rats. The Fig. 6 (a) is for Rat A, and the Fig. 6 (b) is for Rat B. The left two bars in each figure show the distribution of NMSE over 8000 data points after switching to the BC stage. We can see that the average decoding results obtained by GaPP are 74% better than those from DSMCPP in Rat A and 32% better in Rat B. We conduct a pair-wise Student's t-test against the alternative specified by the right tail test $MSE_{DSMCPP} > MSE_{GaPP}$ across 15 task switches. Under the null hypothesis at $\alpha = 0.05$, the probability of observing an equal or higher value in the test statistics is indicated as a p-value. We can see that GaPP reconstructs the kinematics significantly better, with p-values smaller than 0.05.

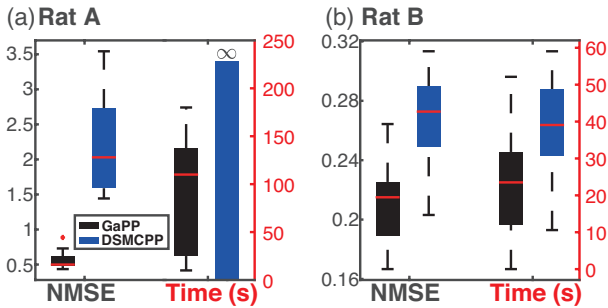

Fig. 6. Statistical performance of kinematics reconstruction of test data of two rats. (a) is for Rat A and (b) is for Rat B. In each figure, the left part shows the mean and variance of normalized mean square error (NMSE) of kinematics reconstruction by GaPP (black bar) and DSMCPP (blue bar) in early stage of BC. And right part shows the mean and variance of the time (red vertical axis) by GaPP (black bar) and DSMCPP (blue bar) when the decoding achieves 90% of the final performance.

The right two bars in each figure compares the distribution of the time when the decoding performance achieves 90% until it finally converges. In Rat A, DSMCPP fails to track the neural tuning direction, and thus the convergence time is set as infinity. Consequently, the kinematics reconstruction by DSMCPP is worse than that of GaPP. In Rat B, the decoding results obtained by GaPP take 43% less time to converge to the final performance. And the p-value is smaller than 0.05 as well against the right tail test $T_{DSMCPP} > T_{GaPP}$ under the null hypothesis at $\alpha = 0.05$. The results indicate that GaPP tracks the neural tuning change more efficiently and results in better kinematics reconstruction.

## IV. DISCUSSION AND CONCLUSION

In this paper, we propose a globally adaptive point process method to track a neural modulation state as the intermediate variable from the discrete spike train and decompose it into the neural tuning parameter prediction and kinematics reconstruction. Neural modulation states measure how the neurons adjust the firing timing in response to different movement. The neural modulation state has the finite range within $[0,1]$, which allows the global search within its full range. The neural modulation state can be further decoupled into the hyper tuning preferred direction without specific boundary by the gradient descent method. In our results, GaPP tracks the 2-dimensional preferred position with more accuracy and less time than DSMCPP, and results in the better decoding performance.

More importantly, the estimated neural modulation state can be directly used to identify neurons that become important in BC. By calculating the mutual information using the spike train and the neural modulation states (modulated kinematics), the neurons with a large amount of information will be selected as those participating most in BC. GaPP tracks the neural modulation state more effectively after switching to BC mode and the result of mutual information demonstrates that the proposed method can identify the neurons becoming more active in BC mode. Using only the important neurons, we can achieve around 90% of the decoding performance of the full recorded neurons. This echoes the findings in [14] where some neurons are selected into the decoding process, the brain selects only a few neurons of them for the adaptation to BC by exploring and altering the tuning properties. Our method provides a computational tool to help understand the efficient mechanism that the brain develops to achieve BC by identifying such neurons and predict how the tuning changes. Moreover, our method is more efficient in identifying the important neurons participating in BC. This is because of the simple calculation of mutual information directly from the neural modulation states. In contrast, the previous methods have to first drive the tuning parameters and reconstruct the kinematics respectively and then calculate MI to measure the importance.

Despite its advantages, our method still has problems that need to be investigated in future. Firstly, our proposed method searches the global optimum for neural modulation states. It helps to identify the important neurons that actively participate in BC. When decomposing the estimated neural modulation states into the hyper tuning parameter, the issue of local minima might emerge in tracking high-dimensional tuning parameter. Thus, for the scenario with high-dimensional parameter such as pairwise interaction among neurons, the advanced decomposition method, such as RMSprop [33], should be considered in future work. Additionally, our method tracks the preferred direction based on the static nonlinear tuning function, and we don't observe a significant change in our data from MC to BC. If the neural data are recorded while the subject is learning a brand-new task, the point process encoding model also needs to take into consideration the change of the tuning nonlinearity.

## REFERENCES

[1] M. S. Gazzaniga, "Plasticity", in *The New Cognitive Neurosciences*. 4th ed., MIT press, 2000, pp. 89-181.

[2] J. P. Donoghue, "Plasticity of adult sensorimotor representations," *Current Opinion in Neurobiology,* vol. 5, no. 6, pp. 749-754, 1995.

[3] C. Schiltz *et al.*, "Neuronal mechanisms of perceptual learning: changes in human brain activity with training in orientation discrimination," *NeuroImage,* vol. 9, no. 1, pp. 46-62, 1999.

[4] I. Mukai, D. Kim, M. Fukunaga, S. Japee, S. Marrett, and L. G. Ungerleider, "Activations in visual and attention-related areas predict and correlate with the degree of perceptual learning," *Journal of Neuroscience,* vol. 27, no. 42, pp. 11401-11411, 2007.

[5] S. Sakellaridi *et al.*, "Intrinsic variable learning for brain-machine interface control by human anterior intraparietal cortex," *Neuron,* vol. 102, no. 3, pp. 694-705.e3, 2019.

[6] F. Gandolfo, C. S. R. Li, B. J. Benda, C. P. Schioppa, and E. Bizzi, "Cortical correlates of learning in monkeys adapting to a new dynamical environment," *Proceedings of the National Academy of Sciences,* vol. 97, no. 5, pp. 2259-2263, 2000.

[7] C.S. Li, C. Padoa-Schioppa, and E. Bizzi, "Neuronal correlates of motor performance and motor learning in the primary motor cortex of monkeys adapting to an external force field," *Neuron,* vol. 30, no. 2, pp. 593-607, 2001.

[8] M. M. Shanechi, "Brain–machine interfaces from motor to mood," *Nature Neuroscience,* vol. 22, no. 10, pp. 1554-1564, 2019.

[9] V. Gilja *et al.*, "A brain machine interface control algorithm designed from a feedback control perspective," in *2012 Annual International Conference of the IEEE Engineering in Medicine and Biology Society*. IEEE, 2012. [Online]. Available: https://dx.doi.org/10.1109/embc.2012.6346180

[10] M. M. Shanechi, Z. M. Williams, G. W. Wornell, R. C. Hu, M. Powers, and E. N. Brown, "A real-time brain-machine interface combining motor target and trajectory intent using an optimal feedback control design," *PLoS ONE,* vol. 8, no. 4, p. e59049, 2013.

[11] S. Dangi, A. L. Orsborn, H. G. Moorman, and J. M. Carmena, "Design and analysis of closed-loop decoder adaptation algorithms for brain-machine interfaces," *Neural computation,* vol. 25, no. 7, pp. 1693-1731, 2013.

[12] J. L. Collinger *et al.*, "High-performance neuroprosthetic control by an individual with tetraplegia," *The Lancet,* vol. 381, no. 9866, pp. 557-564, 2013.

[13] K. Ganguly, D. F. Dimitrov, J. D. Wallis, and J. M. Carmena, "Reversible large-scale modification of cortical networks during neuroprosthetic control," *Nature neuroscience,* vol. 14, no. 5, pp. 662-667, 2011.

[14] A. L. Orsborn, H. G. Moorman, S. A. Overduin, M. M. Shanechi, D. F. Dimitrov, and J. M. Carmena, "Closed-loop decoder adaptation shapes neural plasticity for skillful neuroprosthetic control," *Neuron,* vol. 82, no. 6, pp. 1380-1393, 2014.

[15] W. Truccolo, U. T. Eden, M. R. Fellows, J. P. Donoghue, and E. N. Brown, "A point process framework for relating neural spiking activity to spiking history, neural ensemble, and extrinsic covariate effects," *Journal of Neurophysiology,* vol. 93, no. 2, pp. 1074-1089, 2005.

[16] L. Paninski, "Maximum likelihood estimation of cascade point-process neural encoding models," *Network: Computation in Neural Systems,* vol. 15, no. 4, pp. 243-262, 2004.

[17] Y. Wang and J. C. Principe, "Instantaneous estimation of motor cortical neural encoding for online brain–machine interfaces," *Journal of Neural Engineering,* vol. 7, no. 5, p. 056010, 2010.

[18] E. N. Brown, D. P. Nguyen, L. M. Frank, M. A. Wilson, and V. Solo, "An analysis of neural receptive field plasticity by point process adaptive filtering," *Proceedings of the National Academy of Sciences,* vol. 98, no. 21, pp. 12261-12266, 2001.

[19] L. M. Frank, U. T. Eden, V. Solo, M. A. Wilson, and E. N. Brown, "Contrasting patterns of receptive field plasticity in the hippocampus and the entorhinal cortex: an adaptive filtering approach," *The Journal of Neuroscience,* vol. 22, no. 9, pp. 3817-3830, 2002.

[20] U. T. Eden, L. M. Frank, R. Barbieri, V. Solo, and E. N. Brown, "Dynamic analysis of neural encoding by point process adaptive filtering," *Neural Computation,* vol. 16, no. 5, pp. 971-998, 2004.

[21] A. Ergun, R. Barbieri, U. T. Eden, M. A. Wilson, and E. N. Brown, "Construction of point process adaptive filter algorithms for neural systems using sequential Monte Carlo methods," *IEEE Transactions on Biomedical Engineering,* vol. 54, no. 3, pp. 419-428, 2007.

[22] Y. Wang, A. R. Paiva, J. C. Príncipe, and J. C. Sanchez, "Sequential Monte Carlo point-process estimation of kinematics from neural spiking activity for brain-machine interfaces," *Neural computation,* vol. 21, no. 10, pp. 2894-2930, 2009.

[23] M. M. Shanechi, A. L. Orsborn, and J. M. Carmena, "Robust brain-machine interface design using optimal feedback control modeling and adaptive point process filtering," *PLOS Computational Biology,* vol. 12, no. 4, p. e1004730, 2016.

[24] Y. Wang *et al.*, "Tracking Neural Modulation Depth by Dual Sequential Monte Carlo Estimation on Point Processes for Brain–Machine Interfaces," *IEEE Transactions on Biomedical Engineering,* vol. 63, no. 8, pp. 1728-1741, 2016.

[25] S. Chen, X. Zhang, X. Shen, Y. Huang, and Y. Wang, "Decoding transition between kinematics stages for brain-machine interface*," in *2019 IEEE International Conference on Systems, Man and Cybernetics (SMC)*, 6-9 Oct. 2019, pp. 3592-3597.

[26] W. Q. Malik, W. Truccolo, E. N. Brown, and L. R. Hochberg, "Efficient decoding with steady-state Kalman filter in neural interface systems," vol. 19, no. 1, pp. 25-34, 2011.

[27] B. W. Silverman, "Using kernel density estimates to investigate multimodality," *Journal of the Royal Statistical Society: Series B (Methodological),* vol. 43, no. 1, pp. 97-99, 1981.

[28] J. Chen, K. Xu, Z. Yang, and Y. Wang, "Detecting abrupt change in neuronal tuning via adaptive point process estimation," in *2017 39th Annual International Conference of the IEEE Engineering in Medicine and Biology Society (EMBC)*, 11-15 July. 2017, pp. 4395-4398.

[29] E. N. Brown, R. Barbieri, V. Ventura, R. E. Kass, and L. M. Frank, "The time-rescaling theorem and its application to neural spike train data analysis," *Neural Computation,* vol. 14, no. 2, pp. 325-346, 2002.

[31] Y. Wang, J. C. Principe, and J. C. Sanchez, "Ascertaining neuron importance by information theoretical analysis in motor Brain–Machine Interfaces," vol. 22, no. 5-6, pp. 781-790, 2009.

[32] T. Sharpee, N. C. Rust, and W. Bialek, "Maximally informative dimensions: analyzing neural responses to natural signals," in *Advances in neural information processing systems*, 2003, pp. 277-284.

[32] K. Xu *et al.*, "Local-learning-based neuron selection for grasping gesture prediction in motor brain machine interfaces," *Journal of Neural Engineering,* vol. 10, no. 2, p. 026008, 2013.

[33] G. Hinton, N. Srivastava, and K. Swersky, "Neural networks for machine learning lecture 6a overview of mini-batch gradient descent," *Cited on,* vol. 14, no. 8, p. 2, 2012.